\documentclass[12pt]{article}
\usepackage{latexsym}
\usepackage{amsfonts}
\usepackage{amssymb}
\usepackage{graphicx}
\usepackage[cp1250]{inputenc}

\title{Salpeter equation and causality } 
\author{ Piotr Kosi\'nski\thanks{supported by  University of Lodz grant no. 506/1037.}\\  
Department of Theoretical Physics and Computer Science \\
University of {\L}\'od\'z \\
Pomorska 149/153, 90 - 236 {\L}\'od\'z, Poland.}
\date{}

\begin{document}
\maketitle
\begin{abstract}

Acausal behavior of solutions to free Salpeter equation is considered . It is shown that the formal properties of solutions suggest the acausal
 propagation of quantum phenomena. On the other hand the same properties of solutions describing macroscopic phenomena can be explained without
 appealing to the notion of acausality.

\end{abstract}

\newpage
\section{Introduction}
It is well known that the properties of relativistic spinless wave equation(the Klein-Gordon equation) differ significantly from those of the 
Schroedinger one. The "probability density" entering the continuity equation is not possitive definite; the energy can attain both positive
 and negative values; the so called Klein paradox emerges when scattering in external potential is considered. These paradoxical properties 
can be explained within standard relativistic quantum theory. One has to begin with the quantum scheme sufficiently general to comprise an 
arbitrary number of particles. The next step is to choose the space-time symmetry and implement it by defining the unitary representation
of the group acting in the space of states. Now, the properties of the theory depend strongly on the choice of this group. In the
 nonrelativistic case (Galilei group) it is possible to choose the interaction Hamiltonian which commutes with the particle number
 operator. By selecting the common eigenspace of both operators one arrives at the standard form of N-body nonrelativistic quantum theory.
 On the other hand, the relativistic symmetry (based on Poincare group) seems to imply that there exists no interaction hamiltonian 
commuting with the particle number operator; consequently, the number of particles is not conserved and the N-body sector invariant under 
time evolution cannot be consistently defined (except within some approximation). Important role in drawing the above conclusion is played
 by causality principle which is necessary for relativistic invariance of scattering matrix and implies the existence of 
antiparticles \cite{b1}.

In particular, the one-particle theory is not well-defined or, rather, defined only to some approximation. The properties of relativistic 
wave equations, referred to above, are direct consequences of the structure of relativistic quantum theory. Keeping in mind that any
 measurement is a result of interaction which, under some circumstances can spoil the validity of one-particle approximation one concludes that the problem of the existence of certain
 one-particle observables is highly nontrivial. This concerns, in particular, the position observable which makes the notion of probability
 density in coordinate space questionable.
 
 The above considerations are slightly formal but are supported by more physical arguments based on general properties of quantum theory and
special relativity (see, for example, the beautiful paper \cite{b2} where the penetrating analysis is presented concerning the restrictions
 on measurement accuracy imposed by quantum mechanics in relativistic regime).
 
 The standard scheme sketched above is coherent. There are probably some subtle points which still call for clarification but one can hardly
doubt that the existing paradigm concerning QM and SR cohabitation is correct.

In spite of this state of art there are attempts to formulate the consistent one-particle theory with desired  (i.e. similar to those 
characteristic for nonrelativistic case) properties like positivity of particle energy, clear probabilistic interpretation in coordinate space etc. 
Some of them are based on the so-called Salpeter equation which is basically the square root of Klein-Gordon equation (see the recent paper 
 \cite{b3} and references therein). This is complicated pseudodifferential equation leading to positive definite probability density and,
 from the very construction, to positive energy. However, it has also serious disadvantages. First, it is not manifestly covariant.
 Moreover, being highly nonlocal it can lead to noncausal propagation of particles. Spectral positivity implies some kind of acausal 
behavior even in the case of local dynamics  \cite{b4} $\div$\  \cite{b7}. Within the standard framework it has no serious consequences due to 
the fact that the very notion of localizability loses much of its significance (as compared to nonrelativistic case). However, if one
takes seriously the idea that the relativistic quantum theory admits standard probability interpretation in coordinate space the problem of
(a)causal behavior becomes important.

In the present paper we analyze some simple aspects of acausal behavior of Salpeter equation. Formally, the problem closely resembles 
that considered by Hegerfeldt et al.   \cite{b4} $\div$\ \cite{b7}, i.e. the acausal propagation of positive-energy solutions of
 Klein-Gordon equation. In particular, in second Ref. \cite{b4} Hegerfeldt proved that  even in the case of initial states localized up 
to the exponentially bounded tails the causality (understood as the assumption concerning the finite speed of propagation) is broken.
We present here both some formal arguments (based, in particular, on the results contained in Ref.  \cite{b7}) as well as simple
 intuitive explanation of the phenomena related with propagation described by Salpeter equation.
 
 The paper is organized as follows. Section II is devoted to the study of the simplest case of massless Salpeter equation. Section III 
 deals with the analysis of massive Salpeter equation. Section IV contains some conclusions.

\section{Massless case}
First we consider the free massless particle on a line. The Salpeter equation reads
 \begin{eqnarray}
i\frac{\partial \Phi (x,t)}{\partial t} = \sqrt {-\frac{\partial ^2}{\partial x^2}}\Phi (x,t) ;  \label{1}
\end{eqnarray}

here we adopted the system of units $\hbar =1,\;\; c=1$.\\
Eq.(\ref{1}) implies
 \begin{eqnarray}
\left( \frac{\partial ^2}{\partial t^2} - \frac{\partial ^2}{\partial x^2}\right)\Phi (x,t)=0 \label{2}
\end{eqnarray}
This can be easily seen by differentiating both sides of eq.(\ref{1} ) with respect to time and using again eq.(\ref{1}).
Alternatively, one can use
 \begin{eqnarray}
\pm (i\frac{\partial }{\partial x}) = \sqrt {-\frac{\partial ^2}{\partial x^2}} ,  \label{3}
\end{eqnarray}
which holds locally on the spectrum (not in coordinate space). Concluding, any solution to the Salpeter equation (\ref{1}) solves also (\ref{2}).
However, the inverse is not true. Indeed, eq.(\ref{1}) is a first order evolution equation and its solution is uniquely specified by the initial 
value condition.: $ \Phi (x,t=0)=\Phi _0(x)$. On the other hand, the Cauchy data for eq.(\ref{2}) comprise both $\Phi _0(x)$\ and
 $\dot{\Phi} _0(x)\equiv \frac{\partial \Phi (x,t)}{\partial t}\mid _{t=0}$. One concludes that the solutions to eq.(\ref{1}) are those solutions of
 the wave equation (\ref{2}) for which there exists a specific relation between the initial values for $\Phi (x,t)$\ and
 $\frac{\partial \Phi (x,t)}{\partial t}$. This relation is provided by the spectral positivity condition (see below). Its form is crucial
  in what follows due to the following simple reason. The general solution to eq.(\ref{2}) $\Phi (x,t)=\Psi _1(x-t)+\Psi _2(x+t)$ strongly suggests
  that the wave equation describes propagation with unit velocity. In particular, one expects that if $\Phi (x,0)$\ is nonvanishing only in the 
interval $(-R,R)$, $\Phi (x,t),t\geq 0$, is supported in its causal shadow $\left( -(R+t),(R+t) \right)$. This is, however, true only provided
 $\frac{\partial \Phi (x,t)}{\partial t}\mid _{t=0}$ is also supported in $(-R,R)$\ and, in addition, 
$\int_{-\infty }^{\infty } \frac{\partial \Phi (x,t)}{\partial t}\mid_{t=0}dx=0$. If $ \frac{\partial \Phi (x,t)}{\partial t}\mid_{t=0}$\
 is nonvanishing in some region far outside the interval $(-R,R)$, $\Phi (x,t)$\ will almost immediately (i.e. for small$t>0$) develop nonzero
 value in this region, i.e. outside the causal shadow of $(-R,R)$. So, the question of causal behavior of the solutions to eq.(\ref{1}) reduces 
to the one concerning the supplementary condition which must be imposed on Cauchy data for wave equation (\ref{2}) in order to obtain the
 solution to eq.(\ref{1}): does it imply that $ \frac{\partial \Phi (x,t)}{\partial t}\mid_{t=0}$\ is compactly supported provided $\Phi (x,0)$\
 is ? We shall see that the answer is no.
 
 The solution to the initial value problem defined by eq.(\ref{1}) reads
 
\begin{eqnarray}
&&\Phi (x,t)=\frac{1}{\sqrt{2\pi }}\int_{-\infty }^{\infty } dp e^{i(px-\mid p\mid t)}\tilde{\Phi }_0(p) ,\nonumber  \\
&&\tilde{\Phi }_0(p)= \frac{1}{\sqrt{2\pi} }\int_{-\infty }^{\infty } dx e^{-ipx}\Phi _0(x)   \label{4}
\end{eqnarray}
Now, eq.(\ref{4}) can be rewritten as
\begin{eqnarray}
\Phi (x,t)=\frac{1}{\sqrt{2\pi }}\int_{-\infty }^{\infty } dp \Theta (p) e^{ip(x-t)}\tilde{\Phi }_0(p)+ 
   \frac{1}{\sqrt{2\pi }}\int_{-\infty }^{\infty } dp \Theta (-p) e^{ip(x+t)}\tilde{\Phi }_0(p) \label{5}
\end{eqnarray}
or, using the properties of convolution,
\begin{eqnarray}
\Phi (x,t)=\frac{i}{2\pi }\int_{-\infty }^{\infty } dy \frac{\Phi _0(y)}{(x-t)-y+i\varepsilon }- 
   \frac{i}{2\pi }\int_{-\infty }^{\infty } dy \frac{\Phi _0(y)}{(x+t)-y+i\varepsilon} \label{6}
\end{eqnarray}
Obviously, $\Phi (x,t=0)=\Phi _0(x)$. On the other hand
\begin{eqnarray}
\frac{\partial \Phi (x,t)}{\partial t}=\frac{i}{2\pi }\int_{-\infty }^{\infty } dy \frac{\Phi _0(y)}{\left((x-t)-y+i\varepsilon \right)^2}+ 
   \frac{i}{2\pi }\int_{-\infty }^{\infty } dy \frac{\Phi _0(y)}{\left((x+t)-y+i\varepsilon\right)^2} \label{7}
\end{eqnarray}
or, using elementary properties of distributions
\begin{eqnarray}
\frac{\partial \Phi (x,t)}{\partial t}=\frac{i}{\pi }\int_{-\infty }^{\infty } dy \frac{\Phi _0(y)}{(x-y)^2}     \label{8}
\end{eqnarray}
where the integral is taken in the sense of principal value, i.e.
\begin{eqnarray}
\frac{\partial \Phi (x,t)}{\partial t}=\frac{i}{\pi }\int_{-\infty }^{\infty }\frac{ dz}{z^2}\left(\Phi _0(x+z) + \Phi _0(x-z)-2\Phi _0(x)\right)   \label{9}
\end{eqnarray}
Assume $\Phi _0$\ is supported in the interval $[-R,R]$\ and let $x\gg R$; then $\Phi _0(x+z)=0$, $\Phi _0(x)$\ and 
\begin{eqnarray}
\frac{\partial \Phi (x,t)}{\partial t}=\frac{i}{\pi }\int_{x-R }^{x+R }\frac{ dz}{z^2} \Phi _0(x-z)   \label{10}
\end{eqnarray}
Assuming further $\Phi _0\geq 0$\ we find  $ \frac{\partial \Phi (x,t)}{\partial t}\mid_{t=0}>0$. Therefore, as noted above, $\Phi (x,t)$\ develops nonzero value for small $t$\ in the point outside the causal shadow of $[-R,R]$.
\section{The massive case}
Consider now the massive Salpeter equation
 \begin{eqnarray}
i\frac{\partial \Phi (x,t)}{\partial t} = \sqrt{m^2-\frac{\partial ^2}{\partial x^2}}\Phi (x,t) ,  \label{11}
\end{eqnarray}
together with the initial condition $\Phi (x,t=0)=\Phi _0(x)$. To analyse the (a)causal behavior one can follow the method of Ref. \cite{b7}. 
Assume again that $\Phi _0(x)$\ is smooth and supported in the interval $[-R,R]$\ and $\Phi _0(x)\geq 0$. Since $\Phi _0(x)$\ has a compact support the Paley-Wiener theorem states that its Fourier transform
 \begin{eqnarray}
\tilde{\Phi }_0(p)= \frac{1}{\sqrt{2\pi} }\int_{-\infty }^{\infty } dx e^{-ipx}\Phi _0(x) ,  \label{12}
\end{eqnarray}
is an entire function in complex $p-plane$\ and for any natural $N$\ obeys the estimate
\begin{eqnarray}
\mid \tilde{\Phi }_0(p)\mid \leq  C_N \frac{e^{R\mid Imp\mid }}{(1+\mid p\mid )^N} ,  \label{13}
\end{eqnarray}
Moreover, due to $\Phi _0(x)\geq 0$\
 \begin{eqnarray}
\tilde{\Phi }_0(ip)= \frac{1}{\sqrt{2\pi} }\int_{-\infty }^{\infty } dx e^{px}\Phi _0(x)>0 ,  \label{14}
\end{eqnarray}
Now, the solution to the initial value problem for eq.(\ref{11}) reads
\begin{eqnarray}
\Phi(x,t)= \frac{1}{\sqrt{2\pi} }\int_{-\infty }^{\infty } dx e^{ipx-i\sqrt{p^2+m^2}t}\tilde{\Phi} _0(p) ,  \label{15}
\end{eqnarray}
Let us fix some $t>0$\ and let $x$\ lie outside the causal shadow of $[-R,R]$, say, $x>R+t$. The integrand on the rhs of eq.(\ref{15})
is analytic in the $p-plane$\ with two cuts extending from $-\infty$ to $-im$\ and from $im$\ to $\infty $\ (cf. Fig.1).

\begin{center}
\includegraphics{images-1.eps}\\
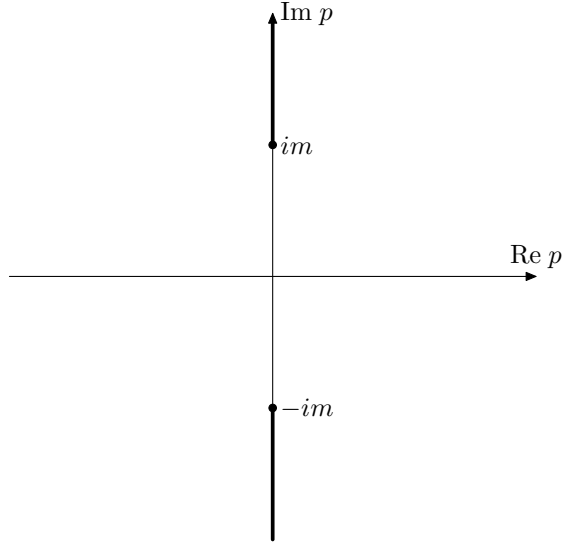
Figure~1.
\end{center}

For $\mid p\mid ^2\gg m^2$\ and any $N$\ one gets, by virtue of eq.(\ref{13}), the estimate

\begin{eqnarray}
\tilde{ \Phi}_0(p)e^{ipx-i\sqrt{p^2+m^2}t}\leq 
 \frac{C_Ne^{(R+t-x)Imp}}{(1+\mid p\mid )^N } ,  \label{16}
\end{eqnarray}
Therefore, the integration contour can be deformed as depicted on Fig.2:
\begin{center}
\includegraphics{images-2.eps}\\
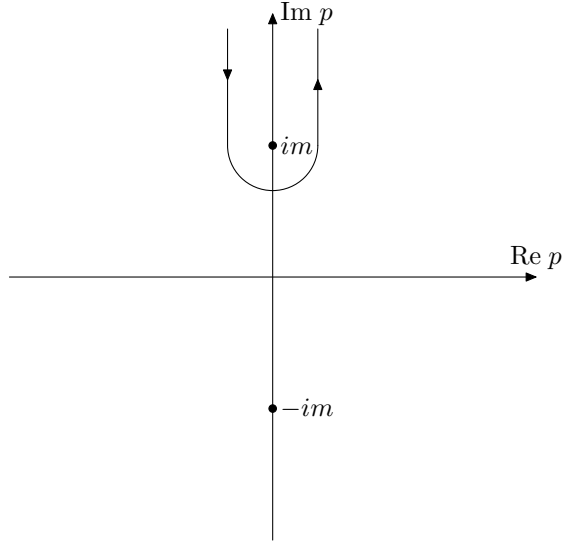
Figure~2.
\end{center}
The integration reduces to that over the discontinuity across the cut. This results
 in the following expression for $\Phi (x,t)$ outside the causal shadow of $[-R,R]$:
\begin{eqnarray}
\Phi(x,t)= i\sqrt{\frac{2}{\pi}}\int_{m }^{\infty } dp\; \tilde{\Phi} _0(ip) e^{-px}sh(\sqrt{p^2+m^2}t) ,  \label{17}
\end{eqnarray}
By virtue of eq. (\ref{14}) the above integral is nonvanishing for $t>0$.
\section {Conclusions}
For better understanding let us reconsider the arguments presented in previous
 sections. To make things simpler we discuss the massless case. The solution to eq.
 (\ref{1}) can be viewed as the solution to the wave equation (\ref{2}) subject to the 
additional condition relating the initial values of the wave function and its time
 derivative. The peculiarity of this condition is that even if the initial wave
 profile is compactly supported the profile of time derivative is not. Therefore,
 infinite tails develop immediately for $t=0^+$.
Alternatively, this phenomenon can be described by inspecting eq.(\ref{6}). Both
 profiles of left- and right- movers are nonlocal and extend over the whole axis.
 For all points $x$, except some finite interval, their values cancel against each
 other. However, for $t>0$\ one profile moves left and the other right so there is no
 cancellation any longer and the resulting wave function extends over all the axis.
	Let us note that no problem with causality arises if our equations describe the
 wave propagation along the material string. For $t>0$\ the nonzero value of wave
 function outside the causal shadow of the support of the initial profile results 
from nonvanishing initial velocity of the corresponding piece of the string and not
 from the “superluminal” propagation of disturbance.

	The situation changes radically in the case of quantum mechanics. Once the
 particle is localized in some domain the reduction postulate states that the
 support of the wave function shrinks to this domain; no other information seems to
 be available. A simple causal explanation of the behavior of wave function is now 
lacking. 
	Recently, the reduction postulate became less popular and there is growing 
conviction that it should be replaced by another idea (for example, decoherence).
 However, it would be not easy to understand the meaning of the above described 
behavior of wave function. It doesn’t seem that the probability interpretation of
 coordinate wave function forces us to make any assumption concerning the value of
 its time derivative outside the domain of particle localization.

 It should be 
 again stressed that, as explained above, the same formal properties of “positive energy” solutions to
 wave equation describing macroscopic phenomena has simple and natural explanations
 having nothing to do with the idea of “superluminal” or acausal propagation.


\begin{thebibliography}{99}
\bibitem{b1}
S. Weinberg, The Quantum Theory of Fields, vol.I, Cambridge University Press, (1995)  
\bibitem{b2}
L. D. Landau, R. Peierls, Zs. Phys. {\bf69}, (1931), 56;
\bibitem{b3}
K. Kowalski, J. Rembielinski, Phys. Rev. {\bf A 84}, (2011), 012108; 
\bibitem{b4}
G. C. Hegerfeldt, Phys. Rev. {\bf D 10}, (1974), 3320; Phys.Rev. Lett. {\bf 54} (1985), 2395;
\bibitem{b5}
G. C. Hegerfeldt, S. N. M. Ruijsenaars, Phys. Rev. {\bf D 22}, (1980), 377;  
\bibitem{b6}
S. N. M. Ruijsenaars, Ann. Phys. {\bf 137}, (1981), 33; 
\bibitem{b7}
P. Kosinski, P. Maslanka, Journ. Math. Phys. {\bf 31}, (1990), 1755.

\end{thebibliography}
\end{document}